\documentclass[11pt]{article}
\usepackage[utf8]{inputenc}
\usepackage{graphicx}
\usepackage{dcolumn}
\usepackage{chngpage}
\usepackage{physics}
\usepackage{float}
\usepackage{amsmath, amssymb}
\usepackage{makecell}
\usepackage{subcaption}
\usepackage{authblk}
\usepackage{indentfirst}
\usepackage{marvosym}
\usepackage{soul}
\usepackage{multirow}
\usepackage{booktabs}
\usepackage[version=4]{mhchem}
\usepackage[
autocite=superscript,
backend=biber,
style=nature,
date=year,
doi=false,isbn=false,url=false,eprint=false
]{biblatex}
\usepackage[lang=en]{jabbrv}
\AtEveryBibitem{\clearlist{language}}
\usepackage[labelfont=bf]{caption}
\usepackage{geometry}
 \usepackage{xurl}
 \usepackage[%
    colorlinks=true,
    linkcolor=blue,
    citecolor=blue,
    urlcolor=black,
]{hyperref}
\usepackage{textcomp, gensymb}
\usepackage{microtype}
\DefineJournalPartialAbbreviation{Rhapsod}{Rhaps}

\title{\textbf{Coherent seeding and control of dynamical ferroelectricity by phonon anharmonicity}}

\author[1,*]{Junhan~Huang}
\author[2,*]{Yongkang~Ju}
\author[3,\Letter]{Xinbo~Wang}
\author[4]{Li~Yue}
\author[1]{Hao~Wang}
\author[1]{Qiaomei~Liu}
\author[1]{Tianchen~Hu}
\author[3,4]{Yuchen~Cui}
\author[1]{Liyu~Shi}
\author[4]{Shangfei~Wu}
\author[3]{Sijie~Zhang}
\author[4]{Dong~Wu}
\author[2,7,\Letter]{Peizhe~Tang}
\author[5,6,1,\Letter]{Tao~Dong}
\author[5,6,1,4,\Letter]{Nan-Lin~Wang}

\affil[1]{International Center for Quantum Materials, School of Physics, Peking University, Beijing 100871, China}
\affil[2]{School of Materials Science and Engineering, Beihang University, Beijing 100191, China}
\affil[3]{Beijing National Laboratory for Condensed Matter Physics, Institute of Physics, Chinese Academy of Sciences, Beijing 100190, China}
\affil[4]{Beijing Academy of Quantum Information Sciences, Beijing 100913, China}
\affil[5]{Tsung-Dao Lee Institute, Shanghai Jiao Tong University, Shanghai 201210, China}
\affil[6]{Zhangjiang Institute for Advanced Study, Shanghai Jiao Tong University, Shanghai 201210, China}
\affil[7]{Max Planck Institute for the Structure and Dynamics of Matter, Luruper Chaussee 149, 22761 Hamburg, Germany}
\affil[*]{These authors contributed equally to this work}
\affil[\Letter]{Email: xinbowang@iphy.ac.cn~(X.W.); peizhet@buaa.edu.cn~(P.T.); taodong@sjtu.edu.cn~(T.D.); nlwang@sjtu.edu.cn~(N.-L.W.)}
\date{}

\begin{document}
\maketitle

\section{Abstract}
Optical control of quantum materials has progressed along two separate directions: creating non-equilibrium states inaccessible at equilibrium, and coherently controlling ultrafast dynamics with multi-pulse protocols.
Ferroelectricity is especially attractive in this context because its order parameter, macroscopic polarization, directly links inversion-symmetry breaking to functional response.
Yet light-induced ferroelectricity has so far been confined to quantum paraelectrics near the ferroelectric instability, where critical fluctuations obscure the formation of a homogeneous ferroelectric state and complicate its deterministic coherent control. 
Unifying these capabilities---preparing a symmetry-broken state and then coherently steering its functionality---remains a central challenge. 
Here we show that intense terahertz excitation of a soft phonon mode induces a ferroelectric state in centrosymmetric PbTe, a thermoelectric material with strong lattice anharmonicity but no ferroelectric transition at finite temperature. The light-induced symmetry-broken state can be realized up to about 100~K, without relying on local dipolar fluctuations.
Experiment and theory together reveal that terahertz-driven anharmonic coupling between degenerate transverse optical phonons underlies this ferroelectric induction.
Furthermore, we demonstrate coherent amplification and suppression of the induced polarization via a double-pulse-excitation protocol.
These results establish terahertz-driven anharmonic mode coupling as a general strategy for controlling mode-mediated functionalities in quantum materials, opening a route to ultrafast information processing.

\section{Main text}

Macroscopic quantum states arise from the minimization of a free-energy landscape shaped by intricate microscopic interactions. Beyond static tuning knobs such as pressure and strain\autocite{Physics_of_quantum_materials_nat_phys2017}, intense ultrafast light fields provide a powerful avenue to dynamically reshape this energy landscape\autocite{bao_peizhe_shuyun_LightinducedEmergentPhenomena2022}, granting access to symmetry-broken states---ranging from charge density waves\autocite{Gedik_Kogar_light-induced_CDW2020,zongRoleEquilibriumFluctuations2021} to magnetism\autocite{Disa_Cava_YTiO3_2023,Disa_Cava_CoF22020,Luo_FePS3_nature2024}---that remain thermodynamically inaccessible in equilibrium. 
In parallel, multi-pulse experiments have shown that light can go beyond merely inducing non-equilibrium states: it can enable on-demand, coherent steering of collective dynamics and phase-switching efficiencies on ultrafast timescales\autocite{coherent_control_huber_Kampfrath_nphoto2011,coherent_control_structual_nature2020,scharfCoherentControlPhonon_Alon_Ron_NC2026a,2DTHz_coherent_control_AFM_kimel_prl2023,maklarCoherentLightControl2023}.
Particularly intriguing in this context is light-induced ferroelectricity\autocite{STO_Li2019,Nova_Cava_STO2019,Li_hQPE_STO2024,KTO_BingCheng2023,Rubio_simu_STO2022}, because it dynamically creates a macroscopic polarization that breaks inversion symmetry.
As ferroelectric (FE) polarization is rooted in collective lattice displacement and directly ties broken symmetry to functionality, inducing and manipulating FE order with light holds unusual promise for studying non-equilibrium dynamics and for realizing ultrafast functional control\autocite{FE_control_nelson_PRL2009,Cava_LNO_FEreversal2017}.

However, realizing and harnessing dynamically induced FE states present intertwined challenges. Thus far, successful experimental demonstrations have remained largely restricted to quantum paraelectrics\autocite{STO_Li2019,Nova_Cava_STO2019,Li_hQPE_STO2024,KTO_BingCheng2023,Rubio_simu_STO2022}. A prime example is SrTiO$_3$ where mesoscopic dipolar fluctuations below 30~K serve as precursors to a FE phase transition, though long-range order is suppressed by nuclear quantum fluctuations\autocite{STO_Li2019,Nova_Cava_STO2019}. Because these systems lie near the FE instability, even subtle perturbations can generate a macroscopic polar state\autocite{STO_isotope_FE_prl1999}. While the pre-existing local dipolar fluctuations facilitate terahertz (THz)-driven inversion-symmetry breaking, the nature of the induced state remains unsettled: recent work has proposed an alternative interpretation in terms of a spatially modulated polar-acoustic phase rather than a homogeneous FE phase\autocite{polarization_density_wave_Orenstein_nphys2025}.
This ambiguity bears directly on coherent control by light: when the induced state remains intertwined with pre-existing critical fluctuations, its microscopic origin is no longer determined solely by a coherently driven long-wavelength collective mode, obscuring a clean basis for deterministic coherent manipulation\autocite{coherent_control_huber_Kampfrath_nphoto2011,coherent_control_structual_nature2020,scharfCoherentControlPhonon_Alon_Ron_NC2026a,2DTHz_coherent_control_AFM_kimel_prl2023,maklarCoherentLightControl2023,johnsonAllopticalSeedingLightinduced_Simon_wall_nat_phys2024}.
The central challenges are therefore to realize an unambiguous light-induced FE state beyond quantum paraelectrics, independent of pre-existing order parameter fluctuations, and to further harness the same coherent lattice dynamics that create it to control the FE polarization on ultrafast timescales.

In this work, we demonstrate coherent seeding and control of dynamical ferroelectricity in PbTe, a rock-salt semiconductor exhibiting giant lattice anharmonicity yet strictly lacking a finite-temperature FE transition.
By resonantly driving the doubly degenerate transverse optical (TO) phonons with intense THz fields, we dynamically stabilize an inversion-symmetry-broken state across a broad temperature range up to 100~K. Systematic experimental measurements, corroborated by theoretical simulations, reveal that this dynamical transition is driven by a critical synergy among phonon anharmonicity, mode degeneracy, coherent THz driving, and suppressed thermal damping. Furthermore, by exploiting this phononic pathway, we employ a double-THz-pump protocol to achieve coherent control of the induced FE polarization.
Our results thus extend the concept of light-induced states beyond the restrictive regime of quantum paraelectrics, and establish a general strategy for engineering ferroic functionalities on demand.

PbTe crystallizes in the $Fm\bar{3}m$ rock-salt structure (Fig.~\ref{Fig_main}a). 
Although the zone-center TO phonon mode associated with an incipient FE instability softens upon cooling\autocite{PbTe_mode_softening_1972}, the system remains robustly paraelectric, with a Curie-Weiss extrapolation placing the putative transition around $-160$~K\autocite{PbTe_neutron_new_mode_prb2012} (Fig.~\ref{Fig_main}b). 
Crucially, however, PbTe is renowned for its giant lattice anharmonicity\autocite{Yue_Chen_PbTe_prl2014}, which gives rise to a range of exotic phenomena.
For instance, strong phonon-phonon scattering at high temperatures significantly suppresses thermal conductivity\autocite{neutron_PbTe_NM_Delaire2011}, establishing PbTe as a leading thermoelectric material\autocite{High-Throughput_thermalelectrics_prx2020}.
Furthermore, large thermally activated anharmonic vibrations lead to the formation of correlated local dipoles only above 100~K---an anomalous phenomenon where local symmetry is lowered upon warming\autocite{Bozin_local_dipole2010,PbTe_correlated_local_dipole_prm2018,Aeppli_hidden_order2020}.
This giant anharmonicity, together with the simple rock-salt phonon manifold that precludes the complex inter-mode couplings found in structurally richer systems\autocite{Rubio_simu_STO2022}, makes PbTe a clean platform for investigating the light induction and control of FE polarization through anharmonic lattice dynamics.
\subsection{Temporal and spectral fingerprints of THz field-induced ferroelectricity in PbTe}
\begin{figure}[H]
    \centering
    \includegraphics[width=\textwidth]{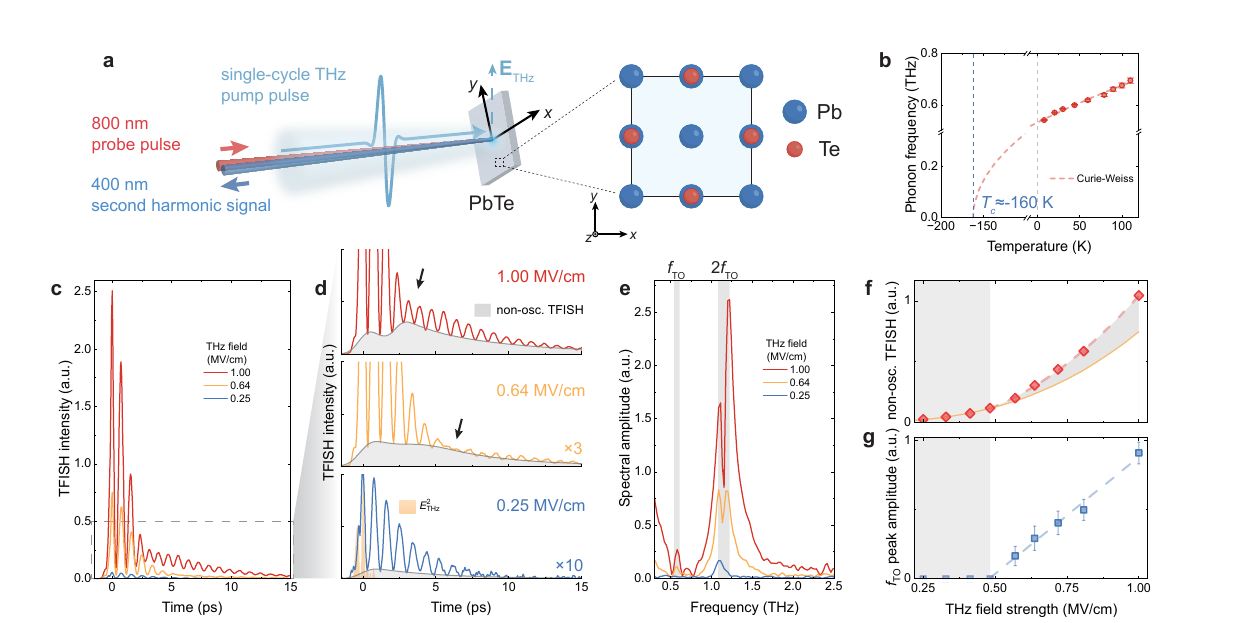}
\caption{
\textbf{Experimental schematics and fingerprints of THz field-induced ferroelectricity.} 
\textbf{a}, Schematic of the TFISH experiment. A single-cycle THz pump pulse (light blue) and an 800~nm probe pulse (red) are focused on the [001] surface of PbTe. The reflected 400~nm second harmonic signal (dark blue) is detected. The probe pulse is polarized parallel to the vertical THz field. Right: Crystal structure of PbTe projected along the [001] direction.
\textbf{b}, Temperature dependence of the TO phonon frequency measured by optical pump and THz probe spectroscopy (Methods). The dashed line shows a Curie-Weiss fit, yielding a negative Curie temperature of about $-160$~K.
\textbf{c}, Representative TFISH signals at 20~K under three different THz field strengths.
\textbf{d}, Zoom-in views of the TFISH signals in \textbf{c}. The shaded gray areas mark the non-oscillatory TFISH component. The shaded yellow area marks the squared profile of the THz pump pulse. Black arrows mark the delayed maximum of non-oscillatory TFISH. 
\textbf{e}, Fourier transforms of the three temporal signals in \textbf{c}. 
\textbf{f}, Field strength dependence of the integrated area of the non-oscillatory TFISH component. The yellow line shows the $\propto E_{\mathrm{THz}}^{2.5}$ fitting for low fields. The dashed line is a guide to the eye.
\textbf{g}, Field strength dependencies of the $f_\mathrm{TO}$ peak amplitude. The shaded gray region marks the range below the threshold field strength for inducing the FE state. Open blue squares represent data points for which the fitting procedure yielded no discernible peak (Methods). The error bars show the standard errors of the fits. Dashed lines are guides to the eye.
}
\label{Fig_main}
\end{figure}

In our experiment, we utilized intense, single-cycle THz pulses to resonantly drive the soft TO mode in PbTe, and measured the second-harmonic generation (SHG) of an 800~nm pulse as a function of the pump-probe time delay (Fig.~\ref{Fig_main}a) to probe the transient inversion symmetry breaking and the ensuing phonon dynamics.
Figure \ref{Fig_main}c,d displays representative traces of THz field-induced second harmonic (TFISH) under selected THz field strengths at 20~K. For the lowest field, the response consists of pronounced oscillations superimposed on a weak, exponentially decaying non-oscillatory background (Fig.~\ref{Fig_main}d, bottom panel). The Fourier transform of these oscillations shows a single peak at twice the TO mode frequency ($2f_\mathrm{TO}$, see Fig.~\ref{Fig_main}e), a signature of homodyne SHG detection of the hyper-Raman active TO phonon\autocite{Cava_YBCO_TFISH_PRX2022,Cava_YBCO_TFISH_npj2025} (Supplementary Note~6). Meanwhile, the non-oscillatory component decays monotonically, and its field dependence follows a quasi-quadratic scaling (Fig.~\ref{Fig_main}f). These responses mirror the behavior of quantum paraelectrics and are attributed to THz-induced local symmetry breaking\autocite{STO_Li2019,KTO_BingCheng2023} (Supplementary Note~2), indicating that no macroscopic FE state is induced in the low-field regime.

Above a threshold field of approximately 0.5~MV/cm, a qualitatively distinct behavior emerges. A delayed local maximum develops in the non-oscillatory component, as marked by arrows in the upper two panels of Fig.~\ref{Fig_main}d. At the highest field, the non-oscillatory signal persists for more than 15~ps, closely resembling the THz field-induced steady state observed in SrTiO$_3$ below 30~K, evidencing the establishment of THz-induced macroscopic FE polarization\autocite{STO_Li2019, Rubio_simu_STO2022, Li_hQPE_STO2024}. 
Notably, the non-oscillatory TFISH intensity deviates significantly from the low-field quasi-quadratic scaling, showing a clear upturn above threshold that signals a strong-field-induced effect (Fig.~\ref{Fig_main}f, see also Fig.~S4a). 
Meanwhile, the establishment of steady state is accompanied by pronounced spectral modifications (Fig.~\ref{Fig_main}e). First, a distinct peak emerges at the fundamental frequency of the TO phonon ($f_\mathrm{TO}$). 
The appearance of the fundamental oscillation in the TFISH measurements originates from heterodyne detection, which strictly requires interference with an auxiliary second harmonic field\autocite{Cava_YBCO_TFISH_npj2025,Cava_YBCO_TFISH_PRX2022}. Since equilibrium PbTe exhibits no SHG (Supplementary Note~5), the auxiliary field acting as the local oscillator must arise from a THz-induced inversion-symmetry-broken state. To the best of our knowledge, this is the first report of a peak at the fundamental frequency in centrosymmetric materials, providing further evidence for the formation of a THz-induced FE state. Second, the $2f_\mathrm{TO}$ peak exhibits asymmetric splitting, arising from renormalization of the driven phonon frequency due to strong anharmonicity, analogous to optical Kerr effects in nonlinear optics\autocite{boyd2008nonlinear} (see also Supplementary Note~4). 
Similar spectral signatures have been observed in nonlinearly driven magnons in orthoferrites, identifying peak splitting as a hallmark of highly nonlinear dynamics\autocite{Huber_PeakSplitting2019,Zhang_nelson_magnon_self-interaction2025}.

Taken together, the concurrent onset of the non-oscillatory TFISH intensity and the $f_\mathrm{TO}$ amplitude above the same threshold field underscores their same origin (Fig.~\ref{Fig_main}f,g), providing compelling evidence for the formation of a THz-induced FE steady state. The accompanying splitting of the $2f_{\mathrm{TO}}$ peak further reveals the strongly anharmonic nature of the lattice dynamics above threshold.

\subsection{Anharmonic behavior of soft phonon}
\begin{figure}[H]
    \centering
   \includegraphics[width=\textwidth]{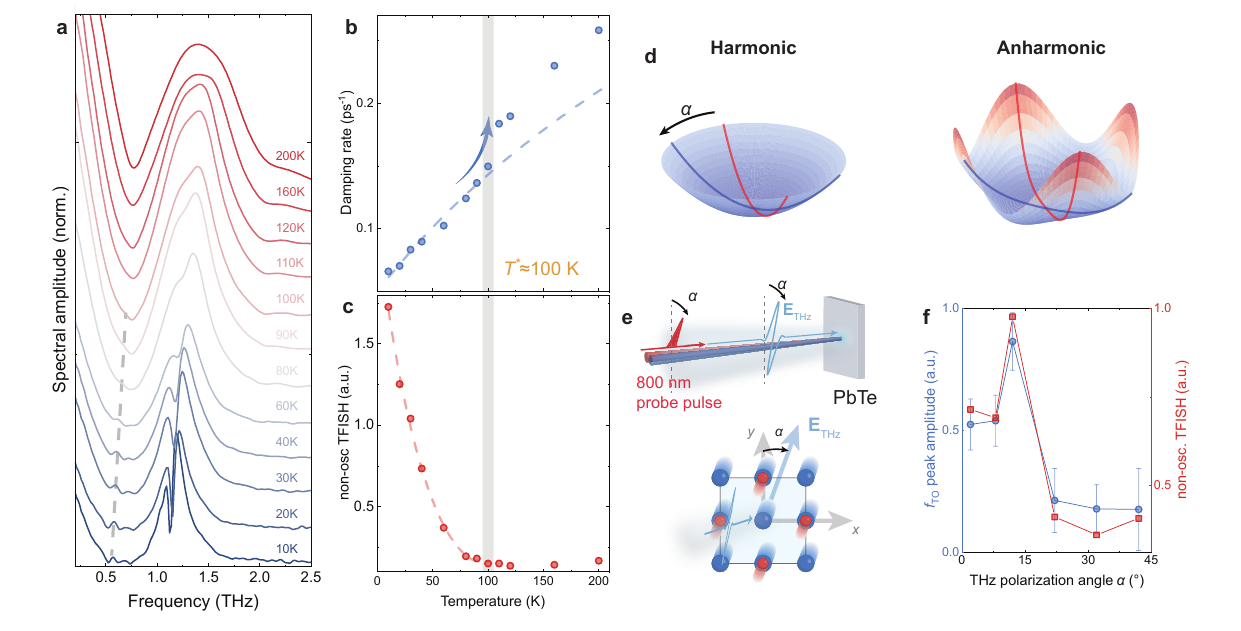}
  \caption{
  \textbf{Anharmonic soft phonon behavior.}
\textbf{a}, Normalized Fourier spectra of the TFISH signal at different temperatures, measured at an excitation THz field strength of 1.00~MV/cm. The dashed line highlights the $f_\mathrm{TO}$ peak.
\textbf{b}, Temperature dependence of the phonon damping rate, measured at the lowest THz field strength of 0.25~MV/cm. The damping rate was extracted from Lorentzian fits. The blue dashed line is a fit to a phonon decay model (see Supplementary Note~8 for details of the extraction and fitting procedures).
\textbf{c}, Temperature dependence of the non-oscillatory TFISH intensity at 1.00~MV/cm, which increases nonlinearly below $T^*\approx 100$~K.
\textbf{d}, Schematic comparison of a harmonic (left) and an anharmonic (right) potential surface for a doubly degenerate phonon. The harmonic potential exhibits continuous rotational symmetry, whereas anharmonicity breaks this rotational invariance.
\textbf{e}, Schematic of the directions of THz polarization and crystal orientation. The THz and probe polarizations were rotated synchronously in the experiment.
\textbf{f}, THz polarization angle $\alpha$ dependence of the $f_\mathrm{TO}$ peak amplitude (blue) and the non-oscillatory TFISH intensity (red).
  } 
  \label{Fig_Anhar}
\end{figure}

Having established a THz field-induced steady state with broken inversion symmetry at low temperature, we next investigated how this dynamical response evolves with temperature and how lattice anharmonicity contributes. 
Figure~\ref{Fig_Anhar}a displays the temperature-dependent Fourier spectra of the TFISH signal under the maximum THz field excitation (1.00~MV/cm).
At low temperatures, the $2f_\mathrm{TO}$ peak is clearly split, consistent with the highly nonlinear phonon dynamics described above. In the high-temperature regime, this splitting is no longer clearly resolved and the peak broadens substantially, indicating that increased damping suppresses the large-amplitude coherent phonon motion required to reach the strongly anharmonic regime.
We therefore extracted the phonon damping rate from the TFISH signals measured at the lowest THz field strength of 0.25~MV/cm (see Supplementary Note~8 for details).
The extracted phonon damping rate (Fig.~\ref{Fig_Anhar}b) grows gradually with temperature, but exhibits a marked upturn above $T^*$, suggesting the onset of additional anharmonic decay channels.

Importantly, $T^*$ also marks the upper temperature limit of the THz-induced FE state. Both the non-oscillatory TFISH intensity (Fig.~\ref{Fig_Anhar}c) and the $f_\mathrm{TO}$ amplitude (Fig.~\ref{Fig_Anhar}a; see also Fig.~S11), which serve as signatures of inversion-symmetry breaking, are prominent only below $T^*$. These observations indicate that the FE state can be induced up to 100~K, significantly higher than $\sim30$~K in SrTiO$_3$\autocite{STO_Li2019}.
Notably, $T^*$ aligns with the previously reported anomaly in PbTe\autocite{Bozin_local_dipole2010,Aeppli_hidden_order2020}, above which thermally activated anharmonicity gives rise to local dipoles\autocite{PbTe_correlated_local_dipole_prm2018} and strongly enhanced phonon scattering\autocite{neutron_PbTe_NM_Delaire2011}. This enhanced phonon scattering is directly reflected in the sharp upturn of the damping rate above $T^*$ (Fig.~\ref{Fig_Anhar}b). Together, these results indicate that the THz-induced FE state in PbTe is governed by coherently driven nonlinear phonon dynamics, rather than by the presence of thermally generated local dipoles. 
Unlike in SrTiO$_3$, where a THz-induced FE state was realized only in the presence of local dipolar fluctuations\autocite{STO_Li2019}, the high-temperature local-dipole regime in PbTe is accompanied by strong dissipation that disrupts the coherent nonlinear dynamics required to establish a macroscopic FE state.

To further investigate the microscopic form of the low-temperature phonon anharmonicity, we performed two-dimensional nonlinear spectroscopy, a powerful approach for resolving nonlinear interactions among collective modes\autocite{Nelson_2D_nphys2024_1,Nelson_2D_nphys2024_2,CoF2_lattice_spin_Kimel_science2021,2DTHz_Jigang_wang_nat_rev_phys2026}. While the nonlinear spectrum reveals clear signatures of highly anharmonic phonon response (Fig.~S18b), the observed features cannot be captured within a single-anharmonic-mode framework (Supplementary Note~13). Instead, a minimal model incorporating the coupling between the two degenerate TO modes suffices to reproduce the experimental features without invoking other phonon branches. This finding suggests that nonlinear phonon response arises from the interplay between the giant anharmonicity and the two-fold degeneracy of the TO phonon.

The anharmonic coupling between the two degenerate modes naturally dictates a polarization-dependent dynamical response to the driving field\autocite{Radaelli_nonlinear_phononics2018,Disa_Cava_CoF22020}. Physically, the strong anharmonicity distorts the two-dimensional potential energy surface away from the harmonic potential (Fig.~\ref{Fig_Anhar}d), thus changing how the THz polarization selects different driving pathways through the anharmonic landscape. As a result, the efficiency for inducing the macroscopic inversion-broken response becomes polarization dependent.
To test this directional control experimentally, we measured the angular dependence of the TFISH response by rotating the THz field polarization, with the angle $\alpha$ defined between the THz polarization and the crystalline $y$ axis (Fig.~\ref{Fig_Anhar}e).
The polarization of the initially vertical THz field was rotated using a pair of wire-grid THz polarizers (WGPs), with the THz field strength fixed at 0.85~MV/cm (Methods).
Measurements were conducted at 20~K. The extracted $f_\mathrm{TO}$ peak amplitude and the non-oscillatory TFISH intensity both show a strong dependence on $\alpha$ (Fig.~\ref{Fig_Anhar}f), reaching a maximum near $\alpha \approx 13^\circ$ and decreasing substantially toward larger angles. This observation suggests that the FE state is more efficiently induced when the THz field is polarized away from the high-symmetry axes.

\subsection{Nonlinear coupling between two degenerate TO modes}
To elucidate the microscopic mechanism by which the nonlinear dynamics of the strongly driven TO phonon generates the FE polarization, we formulated a theoretical model based on first-principles density functional theory (DFT). Figure \ref{Fig_theory}a displays the calculated phonon dispersion of PbTe. Motivated by our experimental observations, we focused solely on the zone-center doubly degenerate TO modes and constructed a minimal anharmonic potential energy surface for them. Because the THz field polarization lies at an angle $\alpha$ with respect to the crystal axes, we defined the normal coordinates $Q_{y'}$ and $Q_{x'}$, in a coordinate frame rotated  by $\alpha$ (Fig.~\ref{Fig_theory}b). Here $Q_{y'}$ represents the mode directly excited by the THz field, while $Q_{x'}$ denotes the orthogonal coupled mode. Constrained by the crystalline symmetry, the anharmonic potential takes the form:
\begin{equation}
V_{\mathrm{ah}}(Q_{x'},Q_{y'})=c_0(\alpha)\left(Q_{x'}^4+Q_{y'}^4\right)+c_1(\alpha)Q_{x'}^2Q_{y'}^2+c_2(\alpha)\left(Q_{x'}^3Q_{y'}-Q_{x'}Q_{y'}^3\right).
\label{eq: DFT_potential_surface}
\end{equation}
Here, the angle-dependent coefficients $c_i(\alpha)$ arise from expressing the anharmonic potential in the rotated coordinate frame $\{Q_{x'},Q_{y'}\}$. Physically, this angular dependence captures how the THz drive selects distinct trajectories on the anharmonic potential.
The coefficients were obtained from DFT calculations (see Methods for the symmetry analysis and computational details).

We then performed dynamical simulations based on this anharmonic potential for a THz polarization oriented at $\alpha=13^\circ$, corresponding to the experimental optimal value (Fig.~\ref{Fig_Anhar}h). Figure \ref{Fig_theory}c displays the simulated time evolutions of the phonon modes $Q_{y'}$ and $Q_{x'}$ alongside the corresponding driving field. 
Initially, the single-cycle THz pulse excites the $Q_{y'}$ mode, driving it to a large amplitude. Through the interactions defined by Eq.~(\ref{eq: DFT_potential_surface}), the motion of $Q_{y'}$ exerts a nonlinear driving force on the orthogonal $Q_{x'}$ mode. Crucially, because $Q_{y'}$ and $Q_{x'}$ are energetically degenerate, this nonlinear driving is resonant, thereby efficiently exciting $Q_{x'}$ into large-amplitude oscillations. The simultaneous substantial excitation of these coupled modes ultimately triggers a dynamical instability within the lattice\autocite{Subedi_nonlinear_phononics2014,Mankowsky_Cava_YBCO_nonlinear_phononics2014}: these phonon modes do not oscillate around their equilibrium positions; instead, both coordinates oscillate around finite quasi-static offsets (yellow lines in Fig.~\ref{Fig_theory}c), signifying the transition into a symmetry-broken state.
The calculated transient FE polarization corresponding to the combined displacements of $Q_{x'}$ and $Q_{y'}$ is shown in Fig.~\ref{Fig_theory}d,e. 
The polarization reaches its maximum about 2~ps after the THz excitation, in qualitative agreement with our experimental observations, with the polarization direction remaining close to $\theta=0^\circ$ (Fig.~\ref{Fig_theory}d). 
This symmetry-breaking distortion, visualized by the real-space displacements of the Pb and Te atoms in Fig.~\ref{Fig_theory}f, lowers the symmetry from the centrosymmetric $m\bar{3}m$ point group to a polar group, consistent with our TFISH polarimetry measurements (Supplementary Note~10).
Furthermore, simulations performed across a range of angles (see Fig.~S17) qualitatively reproduce the observed angular anisotropy of the TFISH response, with the induced polarization reaching its maximum near $\alpha \approx 13^\circ$, in agreement with the experimental maximum in Fig.~\ref{Fig_Anhar}f. The physical origin of this anisotropic FE response lies in the angular dependence of the nonlinear interaction between $Q_{y'}$ and $Q_{x'}$: both the energy transfer efficiency between the two modes and the stability of dynamical trajectories are governed by $c_i(\alpha)$, together making the generation of the quasi-static offset most efficient at intermediate angles.

\subsection{Coherent control of THz-induced FE state}
\begin{figure}[H]
    \centering
   \includegraphics[width=\textwidth]{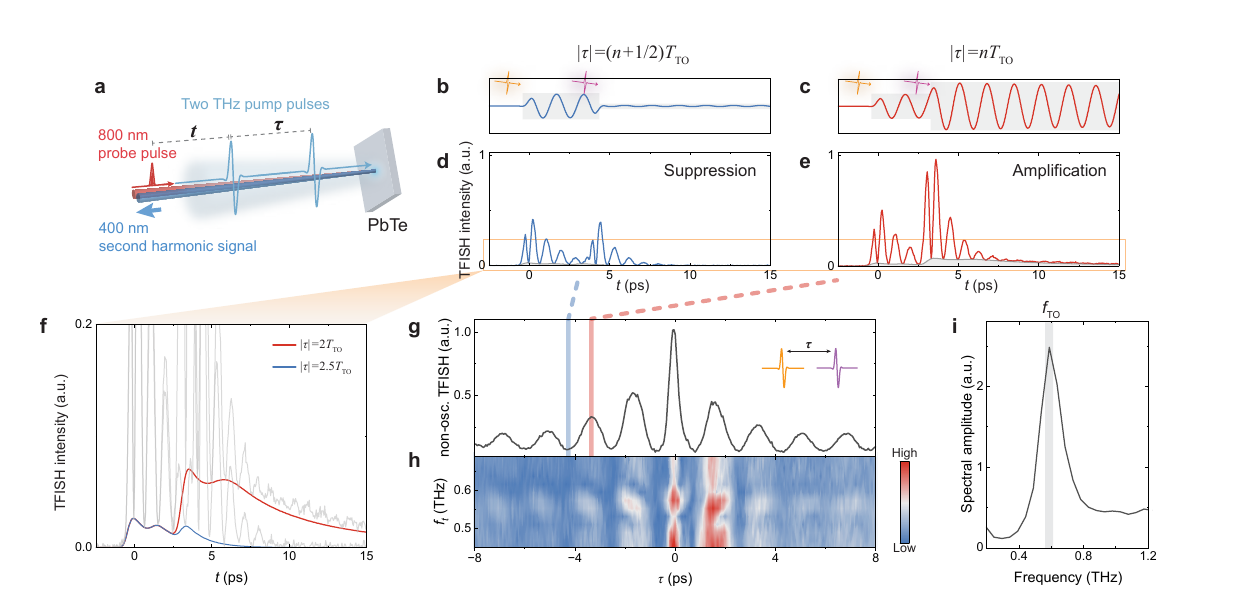}
  \caption{
  \textbf{Coherent control of the FE steady state using a double-THz-pump scheme.}
  \textbf{a}, Schematic of the double-THz-pump experiment. Two parallel-polarized THz pump pulses with a tunable time delay $\tau$ are focused on the sample. The time delay between the THz pulse and the 800~nm probe pulse is denoted as $t$.
  \textbf{b,c}, Schematic illustrations of the evolution of TO phonon amplitude under double-pulse excitation. The TO phonon amplitude can be either suppressed ($|\tau|=(n+1/2)T_\mathrm{TO}$, \textbf{b}) or enhanced ($|\tau|=nT_\mathrm{TO}$, \textbf{c}) by adjusting the delay $\tau$.
  \textbf{d}, TFISH trace obtained at $|\tau|=2.5T_\mathrm{TO}$. The gray shading highlights the non-oscillatory TFISH signal. Upon the arrival of the second pulse, the induced FE state is switched off.
  \textbf{e}, Same as in \textbf{d} but for $|\tau|=2T_\mathrm{TO}$, showing the amplification of the non-oscillatory signal by the second pulse.
  \textbf{f}, Zoom-in views of the TFISH signals in \textbf{d} and \textbf{e}. Blue and red lines correspond to the non-oscillatory components (gray shadings) in \textbf{d} and \textbf{e}, respectively.
  \textbf{g}, Extracted non-oscillatory TFISH intensity as a function of $\tau$.
  \textbf{h}, Fourier transform of the TFISH response with respect to $t$, plotted as a function of $\tau$.
  \textbf{i}, Fourier transform of the oscillations in \textbf{g}, revealing a peak around 0.59~THz, corresponding to the TO phonon frequency $f_\mathrm{TO}$.
  }
  \label{Fig_2D}
\end{figure}

The phononic origin of the THz-induced FE state offers a unique opportunity for its coherent manipulation on ultrafast timescales. To demonstrate this capability, we introduced a second THz pump pulse identical to the first and controlled the relative time delay $\tau$ between the two pulses (Fig.~\ref{Fig_2D}a). To preserve long-lived phonon coherence, the measurements were performed at 10~K, where thermal damping could be minimized. 
The principles of coherent phononic control are illustrated in Fig.~\ref{Fig_2D}b,c. When the second THz pulse arrives out of phase with the TO phonon oscillation launched by the first pulse---at delays $|\tau| = (n+1/2)T_\mathrm{TO}$ where $T_\mathrm{TO}=1/f_\mathrm{TO}$ is the TO phonon period---the phonon amplitude is suppressed through destructive interference. 
Conversely, when the second pulse arrives in phase with the phonon oscillation ($|\tau| = nT_\mathrm{TO}$), the phonon amplitude is constructively enhanced. Consequently, the magnitude of the induced FE polarization can be coherently switched off or amplified by tuning the relative delay between the two THz pulses.
Figure \ref{Fig_2D}d,e exemplifies the experimental realization of this coherent control. Strikingly, when $|\tau|=2.5T_\mathrm{TO}$, the FE steady state is completely switched off following the second THz pulse (Fig.~\ref{Fig_2D}d,f), manifested by the significant suppression of the non-oscillatory TFISH signal. In stark contrast, in-phase excitation ($|\tau|=2T_\mathrm{TO}$) leads to a pronounced amplification of the non-oscillatory signal (Fig.~\ref{Fig_2D}e,f).
Upon continuously varying the delay $\tau$, we monitored two key signatures of the induced FE state: the non-oscillatory TFISH intensity (Fig.~\ref{Fig_2D}g) and the spectral amplitude of the $f_\mathrm{TO}$ peak (Fig.~\ref{Fig_2D}h). Both of them exhibit clear oscillations as a function of $\tau$, demonstrating exceptional control over the non-equilibrium state using a double-pulse-excitation protocol. The Fourier transform (Fig.~\ref{Fig_2D}i) of the oscillations in Fig.~\ref{Fig_2D}g yields a broad peak around the TO phonon frequency, further corroborating the coherent nature of this manipulation.

\subsection{Conclusion and outlook}
In summary, the experiments reported here underscore the power of resonant lattice excitation to induce a FE steady state by exploiting the strong lattice anharmonicity, and demonstrate a strategy to realize and manipulate non-equilibrium functionalities far from equilibrium criticality in high-symmetry crystals.
The microscopic mechanism underlying this non-equilibrium state is the nonlinear coupling between resonantly driven, doubly degenerate TO phonons, whose directionality provides a tuning knob for the magnitude of the induced FE polarization.
This coupling mechanism requires thermal damping to be sufficiently suppressed, in contrast to SrTiO$_3$ where thermal dissipation is required to induce the FE state\autocite{Rubio_simu_STO2022}.
More broadly, coherently driven mode nonlinearities may extend beyond ferroelectricity to engineer a wider range of mode-mediated properties.
Our results also show that strong THz field excitation can serve as a sensitive probe of the equilibrium local hidden order in PbTe\autocite{Aeppli_hidden_order2020}, as manifested by the enhanced phonon damping rate near 100~K, which was previously accessible only through precise neutron measurements\autocite{PbTe_correlated_local_dipole_prm2018,Bozin_local_dipole2010}.
Furthermore, the coherent amplification and suppression of the induced polarization demonstrated here are directly compatible with ultrafast logic operations\autocite{Light-field_control_Boolakee_Hommelhoff2022,coherent_control_valleytronics_Gucci_Cerullo2026}, motivating further exploration for next-generation information processing\autocite{coleCoherentManipulationSemiconductor2001}.

Beyond ferroelectricity, the doubly degenerate zone-center TO mode supports circular ionic motion, forming an axial phonon that carries angular momentum\autocite{Chiral_phonons_Juraschek2025}. At equilibrium, axial phonons in PbTe were shown to carry sizable magnetic moments, measured by the phonon Zeeman effect\autocite{Baydin_Kono_PbTe_chiral_phonon_prl2022}. This suggests that circularly polarized THz fields can, in principle, generate large effective magnetic fields through coherently driven axial TO phonons\autocite{CeF3_chiral_phonon_hanyu2023,STO_dynamical_multi_Basini2024,ultrafast_switching_Barnett_Davies2024}. Our results therefore open new avenues towards dynamically inducing multiple ferroic orders\autocite{Juraschek_dynamical_multi2025}, advancing THz phonon-polaritonics\autocite{baydin2025terahertzcavityphononpolaritons,yaniv2025phononpolaritonhalleffect}, and enabling cavity engineering of quantum materials at THz frequencies\autocite{STO_cavity_Rubio_PNAS2021,cavity_VdW_heterostructure_Kipp2025,Multimode_cavity_Kono_ncomm2025}.

\newpage

\newpage

\newpage
\begin{figure}[H]
    \centering
   \includegraphics[width=\textwidth]{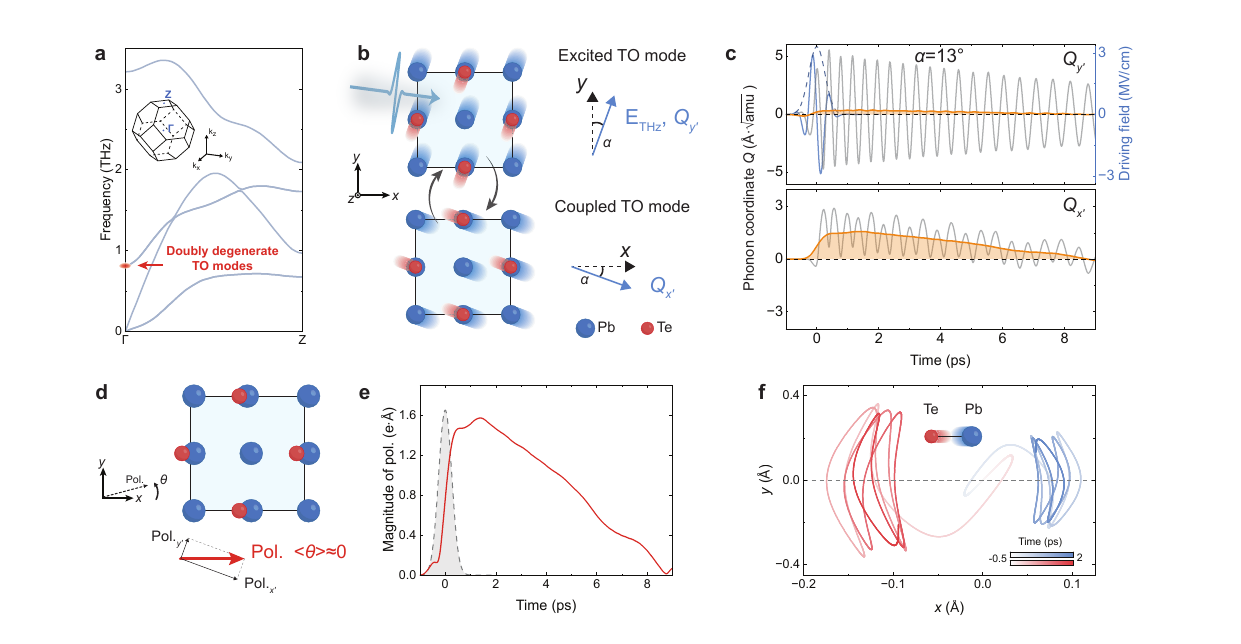}
  \caption{
  \textbf{Anharmonic coupling between the two degenerate TO modes and their dynamics.}
  \textbf{a}, Phonon dispersion of PbTe along the $\Gamma-Z$ direction calculated from DFT. The red circle highlights the doubly degenerate TO modes. Inset: schematic of the Brillouin zone.
  \textbf{b}, Definition of the coordinate systems. The diagram shows the phonon eigenmodes $\{Q_{x'},Q_{y'}\}$ of the doubly degenerate TO phonon, determined by the angle $\alpha$.
  \textbf{c}, Simulated phonon dynamics for the excited mode $Q_{y'}$ (top) and coupled mode $Q_{x'}$ (bottom) for $\alpha=13^\circ$. Solid yellow lines indicate the time-averaged values, revealing steady state offsets highlighted by the yellow shaded regions. Blue solid and dashed lines represent the driving THz field and its envelope used in the simulation, respectively.
  \textbf{d}, Schematic illustration of the induced transient FE polarization, where $\theta$ denotes the angle relative to the crystalline $x$-axis.
  \textbf{e}, Simulated temporal evolution of the FE polarization magnitude. The dashed line shows the THz envelope. A delay between the THz pulse excitation and the peak of induced polarization is observed.
  \textbf{f}, Real-space displacements of Te (red) and Pb (blue) atoms (Methods). The color scales represent the time evolution.
}
  \label{Fig_theory}
\end{figure}

\newpage
\newpage

\section{Methods}

\subsection{Sample preparation}
A stoichiometric mixture of high-purity Pb (99.999\%) and Te (99.999\%) shot (12~g; Pb 7.4265~g, Te 4.5735~g) was loaded into a silica tube, which was evacuated and sealed. The mixture was heated to 1050~\textcelsius, homogenized for 50~h, cooled to 620~\textcelsius\ at 1.5~K/h, and finally cooled to room temperature at 5~K/h. All optical measurements were performed on the [001] surface of the PbTe crystal. The crystallographic $x$ and $y$ axes were determined from the X-ray Laue diffraction pattern.

\subsection{THz pump and SHG probe experiments}
A schematic of the experimental setup is shown in Fig.~S1a. The output of a Ti:sapphire amplifier (35~fs pulse duration, 800~nm central wavelength, 7~mJ pulse energy, 1~kHz repetition rate) was split into two beams. 95\% of the output was used to generate intense single-cycle THz pulses via optical rectification in a MgO-doped LiNbO$_3$ crystal using the tilted-pulse-front technique\autocite{Nelson_tilted_wave2007}. The THz pulses were collimated and focused onto the sample at normal incidence with a set of off-axis parabolic mirrors, yielding a spot size of $\sim$0.55~mm as characterized by a commercial THz camera. The maximum THz field strength at the sample position was estimated to be $1.00$~MV/cm\autocite{Li_hQPE_STO2024,li2025_skyrmion}. The THz field strength was adjusted by a pair of WGPs.
The remaining weak portion of the laser output served as the probe beam to generate the SHG signal from the sample. The probe polarization was controlled by a half-wave plate. The probe pulses were focused onto the sample surface at normal incidence, and the generated 400~nm SHG light was collected in a near-backscattering geometry, filtered by a set of bandpass filters, and detected with a photomultiplier tube (PMT). For the TFISH polarimetry experiment, an additional polarizer was inserted after the probe reflection from the sample. The THz pump beam was modulated at 500 Hz with an optical chopper, and the signal was demodulated at the same frequency using a lock-in amplifier.
All measurements were carried out in a dry air enclosure with the relative humidity maintained below 5\% to suppress absorption of the THz pulses by water vapor. The temporal profile of the THz field, measured by electro-optic sampling in a 1~mm-thick ZnTe crystal, is shown in Fig.~S1b, and the corresponding Fourier transform is shown in Fig.~S1c. 

\subsection{Rotation of the THz field polarization}
The THz field generated via optical rectification in the LiNbO$_3$ crystal using the tilted-pulse-front technique is initially polarized along the vertical direction and cannot be simply rotated by changing the pump polarization. We therefore used a pair of WGPs to rotate it. With the first and second WGPs set at angles $\phi_1$ and $\phi_2=\alpha$, respectively, the transmitted THz field is polarized at the angle $\alpha$, with a field strength of
\begin{equation}
    E_\mathrm{THz}=\cos(\phi_1)\cos(\phi_2-\phi_1)E_\mathrm{max}.
    \label{eq:THz_field_rotation}
\end{equation}
To set the polarization angle $\alpha$, we rotated the second WGP and synchronously adjusted the first WGP so that the field strength remained constant according to Eq.~\ref{eq:THz_field_rotation}. In our experiment, we kept $E_\mathrm{THz}=\cos^2(22.5^\circ)\cdot E_\mathrm{max}\approx0.85E_\mathrm{max}$.
\subsection{Double-THz-pump experiments}
A schematic of the experimental setup is shown in Fig.~S2a\autocite{versatile_2DTHz_setup_wang_RSI2026}. A Ti:sapphire amplifier (35~fs pulse duration, 800~nm central wavelength, 7~mJ pulse energy, 1~kHz repetition rate) served as the primary source. The amplifier had two compressor modules and delivered dual spatially separated outputs, denoted as beam A and beam B. A weak portion of beam A was used as the probe beam. Beam A and the probe beam were each routed through independent translational delay stages, allowing control of the relative time delays between beams A, B, and the probe. Beam B was elevated by 5~cm relative to beam A using a standard periscope configuration incorporating two reflective mirrors.
Two independent THz beams were generated via optical rectification in MgO-doped LiNbO$_3$ crystals using the tilted-pulse-front technique. The generated THz beams were initially polarized along the vertical direction. A $90^{\circ}$ rotated periscope assembly, consisting of a gold-coated reflective mirror and a WGP, was used to switch $E_\mathrm{A}$ into horizontal polarization and then recombine the two THz beams. Consequently, the combined dual THz beams exhibited orthogonal polarization states while preserving their individual electric field strengths.
Another WGP aligned at $45^\circ$ was used to change the two THz beams into a parallel polarization configuration. The THz field strengths of the two beams were both approximately 0.5~MV/cm.
The signal detected by the PMT was sent to a data acquisition card following amplification by a voltage pre-amplifier.
The temporal profiles of the two independent THz fields, measured by electro-optic sampling in a 1~mm-thick ZnTe crystal, are shown in Fig.~S2b, while the corresponding Fourier transforms are shown in Fig.~S2c.

\subsection{Characterization of the \texorpdfstring{$f_\mathrm{TO}$}{fTO} peak amplitude}
For all frequency domain spectra, Fourier transforms were performed using either a Hanning or a Blackman window function with appropriate zero-padding. The $f_\mathrm{TO}$ peak was modeled by a Lorentzian function, $\displaystyle A\left\{[(f-f_0)/(\Gamma/2)]^2+1\right\}^{-1}$, where $A$, $f_0$, and $\Gamma$ denote the peak amplitude, central frequency and damping rate, respectively. To eliminate contributions from the zero-frequency background when characterizing the $f_\mathrm{TO}$ peak, we calculated the second derivative of the Fourier spectrum. The data were then fitted using the analytical second derivative of the Lorentzian function:
\begin{equation}
    A\left\{-\frac{8}{\Gamma^2\left[1+\left(\frac{f-f_0}{\Gamma/2}\right)^2\right]^2}+\frac{128(f-f_0)^2}{\Gamma^4\left[1+\left(\frac{f-f_0}{\Gamma/2}\right)^2\right]^3}\right\}.
\end{equation}
For spectra where the peak was indistinguishable from background noise, defined as a fit yielding a relative standard error $\sigma_A/A > 0.5$, the peak amplitude $A$ was set to zero. Detailed fitting results are presented in Fig.~S4b.

\subsection{Characterization of the TO phonon frequency}
We determined the equilibrium TO phonon frequency using optical pump and THz probe spectroscopy. A 690~nm visible pump pulse was used to generate photocarriers in the sample, and the corresponding change in the low-energy optical conductivity was probed by broadband THz pulses (0.2-3~THz) generated by a photoconductive antenna (Fig.~S19a). This configuration enables detection of infrared-active phonons in the low-frequency range. The pump-induced change in the reflected THz electric field, $\Delta E_{\mathrm{THz}}(t)$, exhibited a coherent oscillation around time zero of the pump-probe delay. The oscillatory component was fitted using $\exp[-\gamma(t-t_0)]\sin(\omega t+\phi)+y_0$ to extract the phonon frequency, as shown in Fig.~S19b. A detailed comparison between the TO phonon frequency obtained in our experiment and previously reported values is provided in Supplementary Note~14.

\subsection{First-principles calculations}
The first-principles calculations of the cubic PbTe crystal were accomplished by using the Vienna ab-initio simulation package\autocite{KresseCMS1996, KressePRB1996} (VASP) based on DFT. A plane-wave basis set combining the projector augmented wave (PAW) method\autocite{BlochlPRB1994} with a kinetic energy cutoff of 250 eV was adopted in our DFT calculations. To describe the exchange-correlation interactions in PbTe, we used the Perdew-Burke-Ernzerhof (PBE) functional\autocite{PerdewPRL1996} based on the generalized gradient approximation (GGA). The primitive unit cell containing one Pb atom and one Te atom was employed, and the corresponding Brillouin zone was sampled with a $10 \times 10 \times 10$ Monkhorst-Pack $k$-grid\autocite{JamesPRB1976}. Lattice parameters and atomic coordinates were fully relaxed until the force imposed on each atom was less than 0.001 {\AA}/eV. Theoretically relaxed lattice parameters are $a = b = c = 4.642$ {\AA}.

Phonon spectrum calculations were performed with the finite differences approach\autocite{GiannozziRMP2001} implemented in the VASP code. A $5\times5\times5$ supercell was employed and sufficiently large to obtain an accurate phonon dispersion. The LO-TO splitting was then introduced by implementing the non-analytical term correction with the Phonopy code\autocite{phonopy-phono3py-JPCM}.

We first focus on the doubly degenerate TO phonon modes at $\Gamma$ point with amplitudes denoted by $Q_x$ and $Q_y$, corresponding to the crystallographic $x$ and $y$ axes, respectively. Constrained by symmetry, the potential takes the form:
\begin{equation}
	V(Q_x, Q_y) = \frac{1}{2}\omega_{0}^2\left(Q_x^2 + Q_y^2\right) + \epsilon\left(Q_x^4 + Q_y^4\right) + \delta Q_x^2Q_y^2,
	\label{eq:potential_surface_without_rotation}
\end{equation}
where $\omega_{0}$ represents the harmonic frequency, while $\epsilon$ and $\delta$ are anharmonic coefficients.
When the sample is rotated by $\alpha$ with respect to the THz polarization, the driven mode $Q_{y'}$ and the nonlinearly coupled mode $Q_{x'}$ are related to the original coordinates $Q_x$ and $Q_y$ through the rotation transformation:
\begin{equation}
\begin{pmatrix}
Q_{x'}\\
Q_{y'}
\end{pmatrix} =\begin{pmatrix}
\cos \alpha  & -\sin \alpha \\
\sin \alpha  & \cos \alpha 
\end{pmatrix}\begin{pmatrix}
Q_{x}\\
Q_{y}
\end{pmatrix}.
\label{eq:rotation_transform}
\end{equation}
By applying the transformation and expressing $Q_x$ and $Q_y$ in terms of $Q_{x'}$ and $Q_{y'}$, we obtained the potential surface in the rotated basis [Eq.~(\ref{eq: DFT_potential_surface})], characterized by the following coefficients: 
\begin{gather}
c_{0}( \alpha ) =\epsilon -\frac{1}{4}( 2\epsilon -\delta )(\sin 2\alpha )^{2},\\
c_1(\alpha)=\delta+\frac32(2\epsilon-\delta)(\sin 2\alpha)^2,\\
c_{2}( \alpha ) =( 2\epsilon -\delta )\cos 2\alpha \sin 2\alpha .
\end{gather}
To determine the coefficients $\epsilon$, $\delta$, and $\omega_{0}$, we performed frozen-phonon calculations by computing the total energies on a two-dimensional grid of atomic displacements along the eigenvectors of the two phonon modes, $Q_x$ and $Q_y$, spanning the range $[-10, 10]$~{\AA}$\sqrt{\mathrm{amu}}$ with a step size of 1.0~{\AA}$\sqrt{\mathrm{amu}}$. The resulting potential energy landscape was then fitted to the model potential [Eq.~(\ref{eq:potential_surface_without_rotation})] to extract the coefficients.
During DFT calculations of the potential surface, the spin-orbit coupling (SOC) effect was taken into account.

The relationship between the real-space atomic displacement $U$ (plotted in Fig.~\ref{Fig_theory}f) and the phonon normal-mode amplitude $Q$ is given by\autocite{Subedi_nonlinear_phononics2014}
\begin{equation}
	\begin{gathered}
	U_{j, \mu}^{\nu} = \frac{Q_{\nu}}{\sqrt{m_j}} \cdot w_{j, \mu}^{\nu}
	\end{gathered}
	\label{eq:normal_mode_amplitudes}
\end{equation}
where $U_{j, \mu}^{\nu}$ denotes the displacement of the $j$-th atom along the Cartesian direction $\mu$ contributed by phonon mode $\nu$. $Q_{\nu}$ is the amplitude of phonon mode $\nu$. $m_j$ is the atomic mass, and $w_{j, \mu}^{\nu}$ is the corresponding component of the phonon eigenvector. The units of $U$, $Q$, and $m_j$ are {\AA}, {\AA}$\sqrt{\mathrm{amu}}$, and amu, respectively, where amu denotes the atomic mass unit. The eigenvector $\vec{w}^{\nu}$ is dimensionless and normalized.
The Born effective charge tensor is calculated using density functional perturbation theory (DFPT)\autocite{GiannozziRMP2001,PhysRevB.55.10355} including the SOC effect. The mode effective charge is given by\autocite{doi:10.1021/acs.jpclett.2c00070}
\begin{equation}
	\begin{gathered}
	Z^*_{\nu, \mu} = \sum_{j} Z^*_{j} \frac{w_{j, \mu}^{\nu}}{\sqrt{m_j}}
	\end{gathered}
	\label{eq:mode_effective_charge}
\end{equation}
where $Z^*_{j}$ is the calculated Born effective charge of atom $j$ and is $\pm8.645e$ for Pb and Te, respectively. Here $Z^*_{j}$ can be taken as a scalar since the Born effective charge tensor is diagonal and isotropic for PbTe by symmetry.
Details on the first-principles calculations and calculations of the phonon potential surface are provided in Supplementary Note~11.

\subsection{Dynamical simulation of nonlinear phonon dynamics}
The classical equations of motion of the two coupled degenerate phonons can be derived from Eq.~(\ref{eq: DFT_potential_surface}) as
\begin{gather}
\ddot{Q}_{y'} +\gamma _{0}\dot{Q}_{y'} +\omega _{0}^{2} Q_{y'} =-4c_0(\alpha)Q_{y'}^3-2c_{1} (\alpha )Q_{x'}^{2} Q_{y'} +3c_{2} (\alpha )Q_{y'}^{2} Q_{x'} -c_{2} (\alpha )Q_{x'}^{3} +F_{\mathrm{THz}}(t), \label{eq:EOM_Qy_expand}\\ 
\ddot{Q}_{x'} +\gamma _{0}\dot{Q}_{x'} +\omega _{0}^{2} Q_{x'} =-4c_0(\alpha)Q_{x'}^3-2c_{1} (\alpha )Q_{y'}^{2} Q_{x'} -3c_{2} (\alpha )Q_{x'}^{2} Q_{y'} +c_{2} (\alpha )Q_{y'}^{3}, \label{eq:EOM_Qx_expand}
\end{gather}
where $\gamma_{0}$ is the damping constant. $F_\mathrm{THz}(t)$ is the external force imposed on the phonon mode $Q_{y'}$ by the driving terahertz electric field, $F_\mathrm{THz}(t)=\vec{Z}_{y'}^*\cdot\vec{E}_{\mathrm{THz}}(t)$, where $\vec{Z}_{y'}^*$ is the mode effective charge and
\begin{equation}
    \vec{E}_{\mathrm{THz}}(t)=E_0\exp\left(-\frac{t^2}{2\sigma^2}\right)\sin(\Omega_{\mathrm{THz}}t+\varphi_\mathrm{THz})\hat{\mathbf{e}}_{y'}
\end{equation}
represents the driving THz electric field. Here only $Q_{y'}$ is directly coupled to the THz electric field. These coupled nonlinear differential equations are solved numerically to obtain the phonon dynamics. Details on the dynamical simulations are provided in Supplementary Note~12.

\section{Data availability}
\noindent The datasets generated and/or analyzed during the current study are available from the corresponding author on request.

\section{Code availability}
\noindent The code used for the current study is available from the corresponding author on request.
\printbibliography

\paragraph{Acknowledgments}
This work was supported by the National Natural Science Foundation of China (Grant Nos. 12488201 to N.-L.W., 12250008 to T.D., 12574349 to X.W., and 12234011 and 12374053 to P.T.); 
and the National Key Research and Development Program of China (Grant Nos. 2024YFA1408700 to N.-L.W., 2025YFA1411504 to T.D., 2024YFA1611300 to X.W., and 2024YFA1409100 to P.T.).
A portion of this work was carried out at the Synergetic Extreme Condition User Facility
(SECUF, \url{https://cstr.cn/31123.02.SECUF}). 
We thank Angel Rubio for fruitful discussions.

\paragraph{Author contributions}
J.H., T.D., and N.-L.W. conceived the study. 
J.H. performed the measurements with help from X.W., H.W., Q.L., L.S., and S.Z.
X.W. developed the experimental setup. 
Y.J. performed the first-principles calculations under the supervision of P.T.
J.H. and Y.J. performed the dynamical simulations under the supervision of P.T.
D.W. and Y.C. synthesized PbTe single crystals.
T.H. and S.W. characterized the samples.
J.H. analyzed the data with help from L.Y. and T.D.
J.H. interpreted the data and wrote the manuscript with input from Y.J., P.T., T.D., and other authors.
T.D., P.T., and N.-L.W. supervised the project.

\paragraph{Competing interests}
The authors declare no competing interests.

\end{document}